\documentclass[useAMS]{mn2e}

\usepackage{graphicx}

\title{
New dynamo pattern revealed by solar helical magnetic fields
}

\author[Hongqi Zhang, et al.]{Hongqi Zhang,$^1$\thanks{E-mail: hzhang@bao.ac.cn.} T.
Sakurai,$^{2}$ A. Pevtsov,$^{3}$ Yu Gao,$^{1}$ Haiqing Xu,$^{1}$
\newauthor
D. D. Sokoloff$^{1,4}$ and K. Kuzanyan,$^{1,5}$\\
$^{1}$National Astronomical Observatories, Chinese Academy of Sciences, Beijing 100012, China,\\
$^{2}$National Astronomical Observatory of Japan, Mitaka 181, Tokyo, Japan,\\
$^{3}$National Solar Observatory, Sunspot, NM 88349, U.S.A.,\\
$^{4}$Department of Physics, Moscow State University, Moscow 119992, Russia,\\
$^{5}$IZMIRAN, Troitsk, Moscow Region 142190, Russia\\}

\date{Accepted ??. Received ??; in original form ??}
\pagerange{\pageref{firstpage}--\pageref{lastpage}} \pubyear{2002}
\def\LaTeX{L\kern-.36em\raise.3ex\hbox{a}\kern-.15em
    T\kern-.1667em\lower.7ex\hbox{E}\kern-.125emX}

\begin{document}






\label{firstpage}

\maketitle


\begin{abstract}
Previously unobservable mirror asymmetry of the solar magnetic field
-- a key ingredient of the dynamo mechanism which is believed to
drive the 11-year activity cycle -- has now been measured. 
This was achieved through systematic monitoring of solar active regions carried out for more than 20 years at observatories in Mees, Huairou, and Mitaka.
In this paper we report on detailed analysis of vector magnetic field data, obtained at Huairou Solar Observing Station in China. 
Electric current helicity (the product of current and magnetic field
component in the same direction) was estimated from the data and a
latitude-time plot of solar helicity during the last two solar
cycles has been produced. We find that like sunspots helicity
patterns propagate equatorwards but unlike sunspot polarity helicity
in each solar hemisphere does not change sign from cycle to cycle -
confirming the theory. There are, however, two significant
time-latitudinal domains in each cycle when the sign does briefly
invert. Our findings shed new light on stellar and planetary dynamos
and has yet to be included in the theory.
\end{abstract}

\begin{keywords}
Sun: activity -- Sun: magnetic fields -- Sun: dynamo.
\end{keywords}

\section{Introduction}

``Whirling storms in the earth's atmosphere, whether cyclones or
tornadoes follow a well-known law which is said to have no
exceptions: the direction of whirl in the Northern hemisphere is
left-handed or anti-clockwise, while in the Southern hemisphere it
is right-handed or clockwise'' (Hale et al. 1919). Eruptive
phenomena in the solar atmosphere are much more powerful than
typhoons in the Earth's atmosphere, and strong helical magnetic
fields in regions where eruptive phenomena occur store huge amounts
of magnetic energy. American astronomer G.E. Hale was the first to
discover the magnetic fields of sunspots and to analyse their
helical magnetic configurations. He recognized a similarity with the
above-cited polarity rule now known as Hale's polarity rule for
sunspots: the sign of magnetic field is anti-symmetric over the
solar equator and changes with every 11-year cycle. Recent
observations of helical structures in the magnetic fields of solar
active regions at the photospheric level encourage us to generalise
Hale' law for the time-latitudinal distribution of the magnetic
field helicity, which plays a key role in the mechanism of magnetic
field generation operating in the solar interior.

In 1955, Parker suggested a dynamo mechanism in the solar interior
based on the combined action of differential rotation and cyclonic
convective vortices (Paker 1955) as a viable way to generate
magnetic fields capable of driving the activity cycle. We are able
to quantify the differential rotation from the motion of large-scale
magnetic fields at the solar surface and from helioseismology in the
solar convection zone. However, because of the opacity of the solar
atmosphere, knowledge of the action of convective vortices can only
be obtained from available observations of helical magnetic fields.
According to mean field dynamo theory, the electromotive force ${\bf
E}$ averaged over convective eddies has a component parallel to the
magnetic field, ${\bf E}=\alpha {\bf B}+...$, where the pseudoscalar
$\alpha$ is related to kinetic and electric current helicities
$\alpha \sim\, <{\bf V}\cdot{\bf {{\bf {\nabla}} \times \,}} {\bf V}> + <{\bf
B}\cdot{\bf {{\bf {\nabla}} \times \,}} {\bf B} >$. The determination of the
kinetic helicity in the solar atmosphere is difficult, while the
twist of magnetic fields, can be estimated from photospheric vector
magnetograms of solar active regions (Seehafer 1990; Abramenko et
al. 1996; Zhang 2006). The formation of twisted magnetic fields
inside the Sun is a fundamental topic for solar dynamo models
(Kleeorin et al. 2003; Choudhuri et al. 2004; Sokoloff et al. 2006;
Zhang et al. 2006); however, for alternative interpretations see
(Tanaka 1991; Zhang 1995). The achievement presented in this Report
is analysis of vector magnetograms of solar active regions, obtained
as a result of systematic monitoring for about 20 years at several
solar observatories, such as at Mees in Hawaii, Huairou in Beijing,
Mitaka in Tokyo, and also Marshall Space Flight Center in
Huntsville. The information on mean electric current helicity and
twist comes from vector magnetograms in the form $H_{{\rm C}z}=({\bf
B}\cdot{\bf {{\bf {\nabla}} \times \,}} {\bf B})_{z}$ or $\alpha_{{\rm ff}}=({\bf
B}\cdot{\bf {{\bf {\nabla}} \times \,}} {\bf B})_{z}/B_{z}^{2}$, where the former
is a part of the current helicity density, and the latter keeps the
same sign as the former one, being a factor of the force-free
magnetic field. These two quantities reflect different helical
characteristics of magnetic fields (see, e.g., {Sokoloff et al.
2008; Kuzanyan et al. 2000 Zhang et al. 2002 for details) and, 
therefore, we consider both of them. 

Obtaining values of these helical parameters averaged over an
active region we may expect that they reflect the handedness of the
solar magnetic field. We found that active regions in the northern
(southern) solar hemisphere possess statistically mainly left
(right) handedness (Seehafer 1990; Pevtsov et al. 1994, 1995; Bao \&
Zhang 1998) which is persistent over the noisy nature of the signal.
Although we admit that the comparison of observational results
obtained at different instruments and their interpretation performed
by various research groups is not a straightforward issue (Pevtsov
et al. 1994; Bao \& Zhang, 1998; Hagino \& Sakurai, 2004), on
average, about 57-66\% of solar active regions follow this, so
called, the hemispheric helicity rule (Pevtsov et al. 1995; Bao \&
Zhang 1998; Hagino \& Sakurai 2004; Bao, Ai \& Zhang 2000; Pevtsov,
Canfield \& Latushko 2001; Abramenko et al. 1997).

\section{Methods}

\begin{figure}
\vspace{7pt} \centering
\includegraphics[width=80mm,angle=0]{
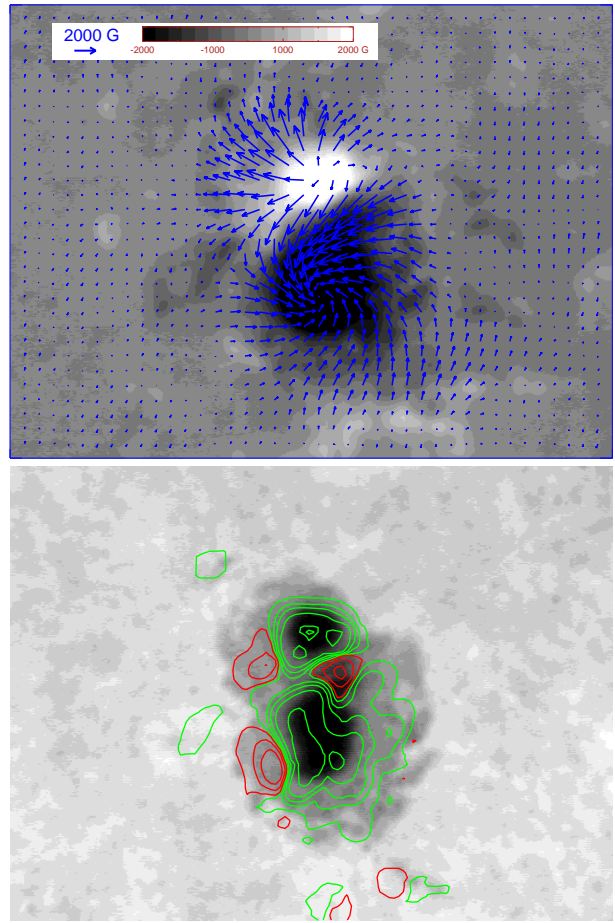}
\caption{
The main part of active region NOAA 6619 taken at Huairou Solar Observing Station on May 11, 1991, 
at 03:26 UT. 
Top: the photospheric vector magnetogram taken. 
Positive/negative values of longitudinal
components of the magnetic field are shaded white/black. The transverse magnetic field is shown by
blue arrows; the magnitude of the field is proportional to the
length of the arrows. Bottom: the contours of electric current
helicity are plotted over the filtergram of this active region;
positive (negative) values are shown by red (green) contours
corresponding to 0.01, 0.05, 0.1, 0.5 G$^{2}$m$^{-1}$, respectively. 
The average values of current helicity and twist are $-8.7\cdot10^{-3}\,{\rm G}^{2}{\rm m}^{-1}$ and $-3.2\cdot10^{-8}\,{\rm m}^{-1}$, the standard deviations $0.08\,{\rm G}^{2}{\rm m}^{-1}$ and $1.1\cdot10^{-6}\,{\rm m}^{-1}$, respectively. One can see that the quantities are highly fluctuating.
The field of view is $2.6^{'}\times1.8^{'}$.
}
\end{figure}

The observational data used in our analysis were obtained at several
observatories in Mees, Mitaka and Huiarou. The magnetographic
instrument at Huairou Solar Observing station is based on the 
FeI $\lambda$5324.19 \AA{} spectral line and determines the magnetic field values at
the photospheric level. The data are obtained from a CCD camera with
512 $\times$ 512 pixels over the whole magnetogram, whose total size
is comparable with the size of an active region. Because of the
observational technique, the line-of-sight field component ${\bf
B}_{z}$ is determined with a much higher precision than the
transverse components (${\bf B}_{x}$ and ${\bf B}_{y}$), where $x$,
$y$, $z$ are local Cartesian coordinates connected with a point on
the solar surface, and the z-axis is normal to the surface. For
details see, e.g. (Abramenko et al. 1996). A measure of mirror
asymmetry of magnetic fields is the electric current helicity
$H_{{\rm C}z}=({\bf B}\cdot{\bf j})_{z}$ , where ${\bf j}=
{\bf {{\bf {\nabla}} \times \,}} {\bf B}$ is the electric current and ${\bf B}$ is the
(small-scale) magnetic field. Because ${\bf {{\bf {\nabla}} \times \,}} {\bf B}$
is calculated from the surface magnetic field distribution, we are
able to determine only the vertical electric current component
$j_{z}=({\bf {{\bf {\nabla}} \times \,}} {\bf B})_{z}$. Therefore, the observable
quantity averaged over an active region is $<H_{{\rm C}z}>=<({\bf
B}\cdot{\bf {{\bf {\nabla}} \times \,}} {\bf B})_{z}> = <B_z j_z>$, see
(Abramenko et al. 1996; Pevtsov et al. 1995; Bao et al. 2000). In the framework of the hypothesis of local homogeneity and isotropy
this value is 1/3 of the current helicity $H_{\rm C}$. In the solar atmosphere conductivity is relatively small and the magnetic field can be locally described as force-free. We consider the force-free factor as another proxy for the mirror asymmetry of the magnetic field which has the meaning of twist. The
twist can be determined as $\alpha_{{\rm ff}}=({\bf B}.{\bf 
{{\bf {\nabla}} \times \,}} {\bf B}) /B^{2}$. 
The observational equivalent of the quantity averaged over an active region 
is the ratio $\alpha_{{\rm av}}=<j_{z}$/$B_{z}>$.
The magnetic field value at each pixel of a magnetogram plays a role of a weighting factor in averages of twist and current helicity over an individual magnetograms. Note, that both the electric current and magnetic field highly fluctuate in space (over the individual magnetogram). One can see that the average of the product and the average of the ratio of these quantities may differ a lot and, in general, not have the same sign. 
%
Note that the signal of current helicity can be detected from solar vector magnetograms with relatively high accuracy rather than one for twist. So, given their different physical meanings, we consider them as two separate characteristics of mirror asymmetry of magnetic field. 

As an example, the photospheric vector field in a typical delta-type solar active region NOAA 6619 
is shown in Fig.~1. One can see that the longitudinal and transverse components of the sunspot magnetic field twist clockwise in the sunspots of the active region (cf. Zhang 2006).

\section{Results}

Fig.~2 shows the distribution of the average helical characteristics
of the magnetic field in solar active regions in the form of
butterfly diagrams (latitude-time) for 1988-2005 (which covers the
most of 22$^{nd}$ and 23$^{rd}$ solar cycles). These results are
inferred from photospheric vector magnetograms recorded at Huairou
Solar Observing Station after statistical reduction of the influence
of magneto-optical (or Faraday) effects in the measurements of
magnetic field (Su \& Zhang 2004; Gao et al. 2008). 
This longest available systematic dataset covering the period of two solar cycles
comprises 6205 vector magnetograms of 984 solar active regions (most
of the large solar active regions of both solar cycles). Of these,
431 active regions belong to the 22$^{nd}$ solar cycle and 553 to
the 23$^{rd}$ one. We have limited the latitudes of active regions
to $\pm$40$^{\circ}$ and most of them are below 35$^{\circ}$ The
helicity values of the active regions have been averaged over
latitude by intervals of 7$^{\circ}$ of solar latitude, and over
overlapping two-year periods of time (i.e., 1988-1989, 1989-1990,
..., 2004-2005). By this way of averaging we were able to group at
least 30 data points in order to make error bars (computed as 95\%
confidence intervals) reasonably small. In this sampling we find
that 66\% (63\%) of active regions have negative (positive) mean
current helicity in the northern (southern) hemisphere over the
22$^{nd}$ solar cycle and 58$\%$ (57\%) in the 23$^{rd}$ solar
cycle.

\begin{figure}
\centering
\hbox{
\hskip -8mm
\includegraphics[width=96mm,angle=90]{
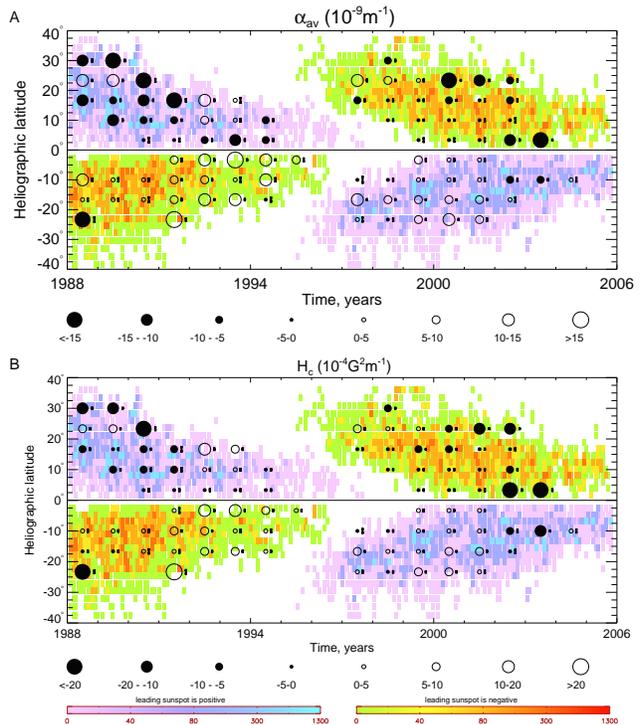}
}
\caption{ Top: the distribution of the averaged twist $\alpha_{\rm
ff}$; and bottom: electric current helicity $H_{{\rm C}z}$ of solar
active regions in the 22$^{nd}$ and 23$^{rd}$ solar cycles.
Superimposed, the underlying coloured ``butterfly diagram'' shows
how sunspot density varies with latitude over the solar cycle.
Vertical axis gives the latitude and the horizontal gives the time
in years. The circle sizes give the magnitude of the displayed quantity. 
The bars to the right of the circles show the level of error bars computed as 99\% 
confidence intervals, scaled to the same units as the circles.
72 out of 88 groups for current helicity (82\%) as well as 67 out of 88 groups for twist (74\%) have the error bars lower than the signal level.
}
\end{figure}

We note a remarkable similarity between the wings plotted for the
different tracers (helicity and twist) in Fig.~2. The analysis of
the data for individual tracers shows some similarity between these
tracers (Sokoloff et al. 2008; Kuzanyan et al. 2000) as well as
between the data obtained at different observatories (Huairou,
Mitaka and Mees) (Hagino \& Sakurai 2005; Pevtsov et al. 2008; Xu et
al. 2007). However, some discrepancy reflects the noisy nature of
the mirror asymmetry, also obtained in direct numerical simulations
(Brandenburg \& Sokoloff 2002; Otmianowska-Mazur et al. 2006).

The message which we infer from this butterfly diagram is as follows:

\begin{enumerate}
 \item Electric current helicity and twist follow the propagation of the magnetic activity dynamo waves recorded by sunspots. This demonstrates that the mirror asymmetry is closely related to the dynamo process. Traditionally, helical motions were considered as necessary for breaking mirror symmetry in dynamos. However, the symmetry can also be broken by random motions (Pevtsov \& Longcope, 2007). The hemispheric helicity rule, shown in Fig.~2, suggests that the solar dynamo is helical. Yeates et al. (2008) have recently computed the pattern of current helicity based on modeling using longitudinal magnetic field data for the 30 months period 1997-1999 which demostrates the hemispheric sign rule inversion at latitudes higher than 50$^\circ$. We can not quantitatively compare our observational results with their computation as there are no reliable observations of active regions at high latitudes. Furthermore, for those years of the solar minimum activity there were no sufficient observational data for statistically significant conclusion.
\item The helicity and twist oscillate with 11-year period like sunspots rather than 22-year period as magnetic fields do. This is strikingly important for the theory, as both quantities are not exactly quadratic in magnetic field.
Here we would like to note that the average amplitude of helicity does not show any significant dependence on the phase of solar cycle. In generating the magnetic field, dynamo flows constantly twist and shear the seed magnetic field. As this magnetic field becomes more twisted, it exerts stronger resistance to further twist. The twist accumulated in a dynamo region may slow-down the dynamo action. This effect is known in dynamo theory as quenching.
 \item The helicity and twist patterns are in general anti-symmetric with respect to the solar equator. This result confirms the hemispheric rule obtained in studies of 11-year observational data sets (Kleeorin et al. 2003; Pevtsov et al. 1994, 1995; Bao \& Zhang 1998).
 \item The helicity pattern is more complicated than Hale's polarity law for sunspots. Our results revealed specific latitudes and times on the butterfly diagram where the hemispheric helicity law is inverted. So, we found areas of the ``wrong'' sign at the end of the butterfly wings. This is a challenge for dynamo theory. We can interpret this phenomenon as penetration of the activity wave from one hemisphere into the other ``wrong'' hemisphere. An analogous pattern can be recognized in sunspot data at the end of the Maunder minimum (Sokoloff 2004). The other domain of the ``wrong'' helicity sign located just at the beginning of the wing has been predicted (Choudhuri et al. 2004) as a result of additional twisting of the magnetic tubes arising to form a sunspot group.
 \item The average amplitude of the helicity does not show any significant dependence on the phase of solar cycle. In generating the magnetic field, dynamo flows constantly twist and shear the seed magnetic field. As the magnetic field becomes more twisted, it exerts stronger resistance to further twist. Too much twist accumulated in a dynamo region may slow-down the dynamo action, the effect is known in dynamo theory as quenching. The lack of a systematic change in the amplitude of helicity of the solar magnetic field over the course of the solar cycle suggests that helicity is continuously removed from the dynamo region.
 \item There is an approximately two year time lag between the sunspots and current helicity and twist patterns: the helicity and twist patterns come after the sunspot pattern. Moreover, the maximum value of helicity, at the surface at least, seems to occur near the edges of the butterfly diagram of sunspots. This is an unexpected result which poses another challenge for dynamo theory. The theory predicts (Parker 1955) a lag of the opposite sign (helicity and twist pattern should appear some 2.7 year before the sunspot pattern). Conventional dynamo models, however, ignore the fact that helicity needs time to follow the magnetic field.
\end{enumerate}

\section{Conclusions}

Let us summarize the results. The mirror asymmetry of magnetic
fields at the solar surface is related with the magnetic field
generation inside the Sun. It evolves with solar cycle, in particular, it has domains of the
``wrong'' helicity sign at the beginning and end of the wings (see Fig.~2). 
A quantitative description of the phenomenon remains for further theoretical studies.

Our results bring up a new type of characteristic of hydromagnetic dynamos and pose a challenge for the theory of stellar and planetary magnetic fields.

\section*{Acknowledgments}
This study is supported by a government project and also grants
10233050, 10228307, 10311120115 and 10473016 of National Natural
Science Foundation of China (NNSFC), and TG 2000078401 and
2006CB806301 of National Basic Research Program of China, as well as
by the cooperative programs between National Astronomical
Observatories of China and National Astronomical Observatory of
Japan and between the NNSFC and Russian Fund for Basic Research
(RFBR) under grant 08-02-92211 and RFBR grants 07-02-00246 and
07-02-00127. Hinode is a Japanese mission developed and launched by
ISAS/JAXA, with NAOJ as domestic partner and NASA and STFC (UK) as
international partners. It is operated by these agencies in
co-operation with ESA and NSC (Norway).  The National Solar
Observatory is operated by the Association of Universities for
Research in Astronomy, AURA, Inc., under a cooperative agreement
with the National Science Foundation.
The authors are thankful to the anonymous referee for constructive criticism which helped improving the Letter.

\end{document}